\documentclass{article}
\usepackage[T1]{fontenc}
\usepackage[latin9]{inputenc}
\usepackage{color}
\usepackage{graphicx}
\PassOptionsToPackage{normalem}{ulem}
\usepackage{ulem}
\usepackage[unicode=true]
 {hyperref}
\usepackage{breakurl}
\begin{document}

\title{\textbf{Visualizing Bags of Vectors}}

\author{Sriramkumar Balasubramanian\hspace{20bp}Raghuram Reddy Nagireddy\\
Computer Science\hspace{115bp}Mathematics\\
\textbf{\normalsize sb3457@columbia.edu\hspace{75bp}rrn2111@columbia.edu}}

\date{May 9th, 2013}
\maketitle
\begin{abstract}
The motivation of this paper is two-fold - a) to compare between two
different modes of visualizing data that exists in a bag of vectors
format b) to propose a theoretical model that supports a new mode
of visualizing data. Visualizing high dimensional data can be achieved
using Minimum Volume Embedding, but the data has to exist in a format
suitable for computing similarities while preserving local distances.
This paper compares the visualization between two methods of representing
data and also proposes a new method providing sample visualizations
for that method.
\end{abstract}

\section{Introduction}

Visualizing data which is in the form of a bag of vectors is useful
in various domains in identifying and understanding relationship between
entities involved. The goal of the visualization is to help the user
identify similarities and differences between data by means of a visual
aid (usually a graph where nodes depict entities involved and the
edges correspond to relationship between the entites). But data belonging
to different domains have different characteristics. Hence a targeted
approach is needed to visualize data for different domains. In this
paper the focus is on a set of authors each of whom has a set of papers
published. \\
\\
The data used for visualization in this paper consists of a set of
authors and their publications. Visualization results are provided
for different author-paper datasets which are defined in \textbf{Sec.
\ref{sec:Experimental-Results}}. \\
\\
Each author can be represented as bag of vectors, where each vector
corresponds to a paper published by that author represented as a vector
of word counts in that document. But this representation of the data
would lead to high dimensional data points to be visualized since
the average number of distinct words in any scientific paper is quite
\textit{large}. Hence dimensionality reduction techniques are necessary
and one such technique is \textbf{MVE }( Minimum Volume Embedding)
\textbf{{[}4{]}}, which is used for the visualization of the data,
in this paper.\\
\\
\textbf{MVE} is an algorithm that facilitates non-linear dimensionality
reduction of data. The reduction consists of \textbf{SDP }(Semi-definite
Programming) and matrix factorization to find a low-dimensional embedding
that preserves local distances between points while representing the
dataset in many fewer dimensions. It has been shown in that MVE improves
upon \textbf{SDE }(Semi-definite embedding), which is the pre-existing
technique, in terms of the resulting eigenspectrum producing better
visualizations for different datasets.\\
\\
While dimensionality reduction of the data is crucial, the representation
of the data provided as an input to \textbf{MVE }mechanism, is equally
important. Usually datasets have to be preprocessed to be presented
in a form \textit{conducive }to visualization. Since \textbf{MVE }entails
techniques to preserve local distances or similarities between the
data, transforming the original dataset to help calculate the similarity
between different data points will aid the visualization process.
One such method of computing similarity scores is using a Bhattarcharya
Kernel \textbf{{[}6{]}{[}7{]} }(based on the Bhattacharya distance)
which is used to compute similarities between probability distributions.
The vectors of word counts for each author can be converted to a vector
of word probabilites by normalization, making it suitable for applying
the Bhattacharya kernel. Also probability distributions can be inferred
by using generative models using the author and paper information
which is the main focus of this paper. \\
\\
This paper is organized as follows :- \textbf{Sec. \ref{sec:Bag-of-Words}.
}discusses the representation of each author as a single vector based
on word counts aggregated over all the papers published by that author.
This is followed by \textbf{Sec. \ref{sec:Generative-Models-for}.
}which introduces the concept of generative models, especially the
\textbf{A-T Model }that is used to infer the underlying probability
distributions which can be used to represent the data. In \textbf{Sec.
\ref{sec:Experimental-Results}. }the actual datasets as well as the
visualization results using the above mentioned representations are
presented. In \textbf{Sec. \ref{sec:Hierarchical-Author-Topic-Model}.
}an extension to the A-T model is presented. This is followed by summing
up the paper with \textbf{Conclusion }and \textbf{Future Work }sections.

\section{Aggregated Bag of Words Model\label{sec:Bag-of-Words}}

In the author-paper dataset, for each author, a set of papers published
by that author constitute the initial dataset. Each paper can be represented
as a vector of word counts. Each dimension of this vector represents
a unique word in the dictionary consisting of unique words from all
the papers of all authors. \textbf{eg. }A paper $p_{ij}$ written
by author $a_{i}$ is represented as $p_{ij}=(c_{ij}^{1},c_{ij}^{2},...,c_{ij}^{|V|})$,
where $c_{ij}^{k}$ denotes the frequency of $k^{th}$ word in the
$j^{th}$ paper author $a_{i}$ written by in the Vocabulary of all
words \textbf{V, }whose size is denoted by $|V|$. We can arrive at
an aggregated word count for each author by aggregating word frequency
vectors for each paper written by that author, \textbf{i.e. }$a_{i}=(c_{i}^{1},c_{i}^{2},....,c_{i}^{|V|})$
where $c_{i}^{k}=\sum_{l=1}^{|P_{i}|}c_{il}^{k}$ , where $P_{i}$
represents the set of papers under author $a_{i}$. But using just
the word counts cannot appropriately capture the similarities between
authors, whereas a normalized vector might be more suitable for comparison
since the number of words in a paper is not constant. This transforms
the aggregated word frequency vector to a probability vector thus
rendering the use of the Bhattacharya Kernel possible. Now $a_{i}^{'}=(p_{i}^{1},p_{i}^{2},...,p_{i}^{|V|})$
where $p_{i}^{k}=c_{i}^{k}/(\sum_{l=1}^{|V|}c_{i}^{l})$. This is
the transformation used on the inital dataset to be used for visualization
using \textbf{MVE. }

\section{Generative Models for Documents\label{sec:Generative-Models-for}}

Generative Models are used to infer the underlying joint probability
distribution of variables defined by graphical models, from which
the conditionals can be derived. These conditionals can be used to
determine an expression like $p(features|a_{j})\forall j$ where $a_{j}$
refers to the $j^{th}$ author, and \textit{features} represents the
characteristics of the data.

\subsection{LDA}

Latent Dirichlet Allocation (\textbf{LDA){[}1{]} }is used to infer
the distribution of the data in terms of a set of hidden topics or
categories. Using the generative paradigm, $p(c_{i}|d_{j})\forall i,j$
can be inferred, where $c_{i}$ refers to the $i^{th}$ hidden category
and $d_{j}$ refers to the $j^{th}$ document. Here we do not use
the authorship information, hence there is no basis for comparison
between authors by inferring this probability distribution as it is
more of a relation between a vector of words over another vector of
words. This problem gets magnified when there are multiple authors
per document. This intuition is further supported by the visualization
depicted in \textbf{Sec. \ref{sec:Experimental-Results}.}

\subsection{Author-Topic (A-T) Model}

The A-T model\textbf{{[}8{]}} builds on \textbf{LDA} by including
the authorship information in influencing the topics assigned to words.
Since the authorship information is also included in the joint of
the entire model, we can arrive at the topic distributions for each
author - $c_{ij}=p(t_{i}|a_{j})\forall i,j$ where $t_{i}$ refers
to the $i^{th}$ hidden topic and $a_{j}$ refers to the $j^{th}$
author. Thus each author can be represented by his/her topic distribution
vector which can now be used to compute similarities and provided
as input for the \textbf{MVE. }

\section{Experiments\label{sec:Experimental-Results}}

The main goal of the experiments is to compare the quality of the
visualization between the \textbf{Aggregated Bag of Words Model }and
the \textbf{A-T Model}. Multiple datasets are used to faciliate the
comparison. The A-T model was derived from the open source \textbf{Topic-Toobox
}software \textbf{{[}2{]}. }Apart from the comparison, experiments
were conducted to generate visualizations on a new model proposed
in \textbf{Sec. \ref{sec:Hierarchical-Author-Topic-Model}. }

\subsection{Dataset}

The datasets used for the visualization experiments include :-
\begin{itemize}
\item A set of 106 authors with 5 paper abstracts each from the set of members
listed at \textcolor{blue}{\uline{\href{http://idse.columbia.edu/people/all}{http://idse.columbia.edu/people/all}}}
\item A set of 68 authors with 10 full papers each from the same source
as above 
\end{itemize}
The datasets were collected manually from \textcolor{blue}{\uline{\href{http://scholar.google.com}{http://scholar.google.com}}}\textcolor{black}{{}
in the form of pdf's and converted to word counts using PDFBox - an
open-source PDF to TXT converter.}

\subsection{Observation}

The visualizations are provided in the form of figures. In some figures
the entire network of connected authors is provided for the reader
to compare and contrast with:-\\
\textbf{}\\
\textbf{Note: }Please magnify the document (electronic version) to
observe the ``subnetwork'' relationships with greater clarity.\\
\\
The original picture files are provided along with the report as well
in .eps format.\\
\\
The fidelity scores for all the visualizations were in the range {[}0.97-0.99{]}.\\
\includegraphics[scale=0.6]{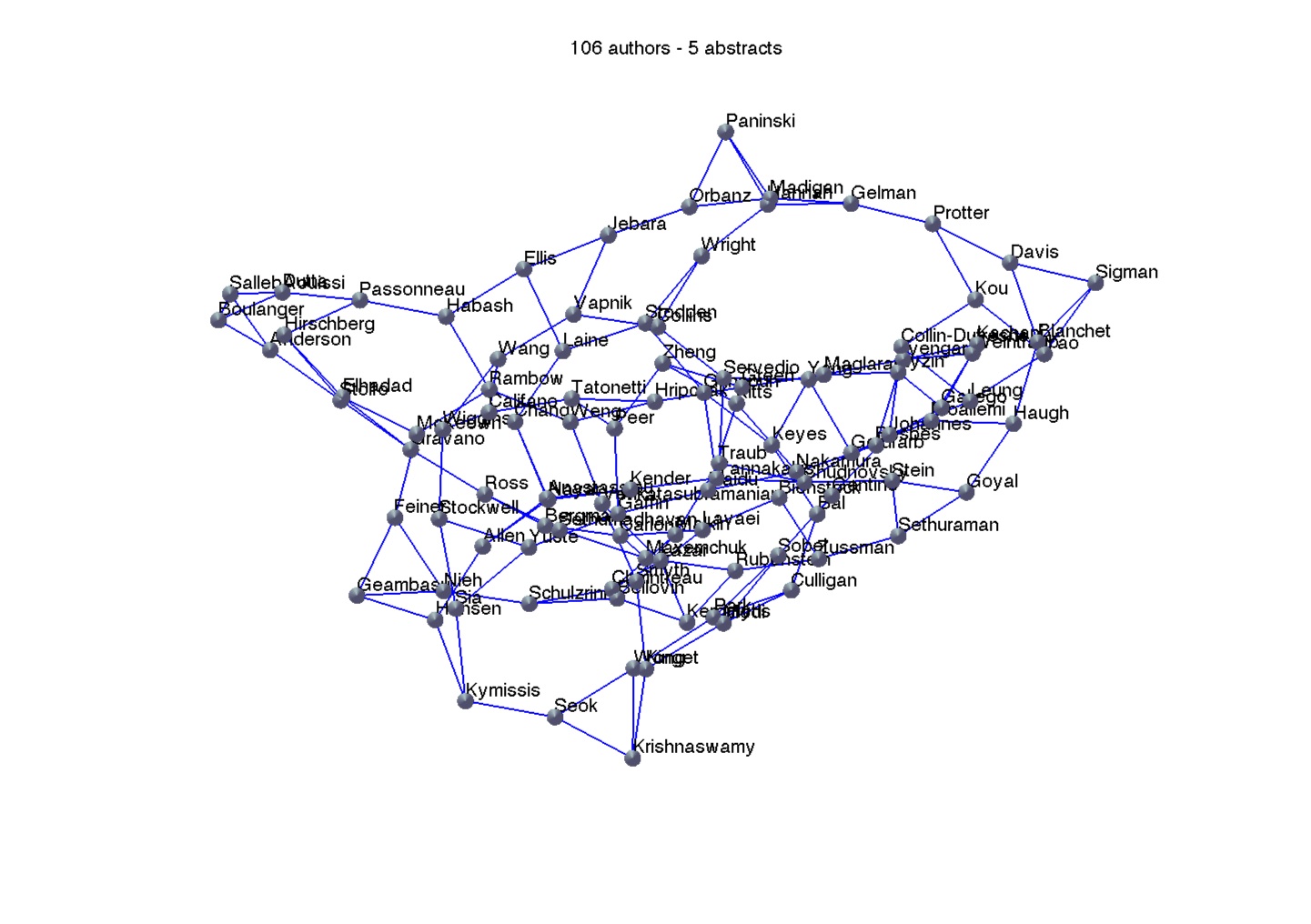}

\begin{center}
\textit{\footnotesize Figure 1. Visualization of Aggregated Bag Of
Words Model - 106 authors - 5 abstracts} 
\par\end{center}

The above figure shows the visualization using the aggregated bag
of words model on the abstracts. There are a few relationships which
are captured, as shown in the diagram below, which tally with the
expected behavior. However apart from a few such ``subnetworks''
there are many deviations from expected behavior that were observed
as well.\\
\\
\includegraphics[scale=0.58]{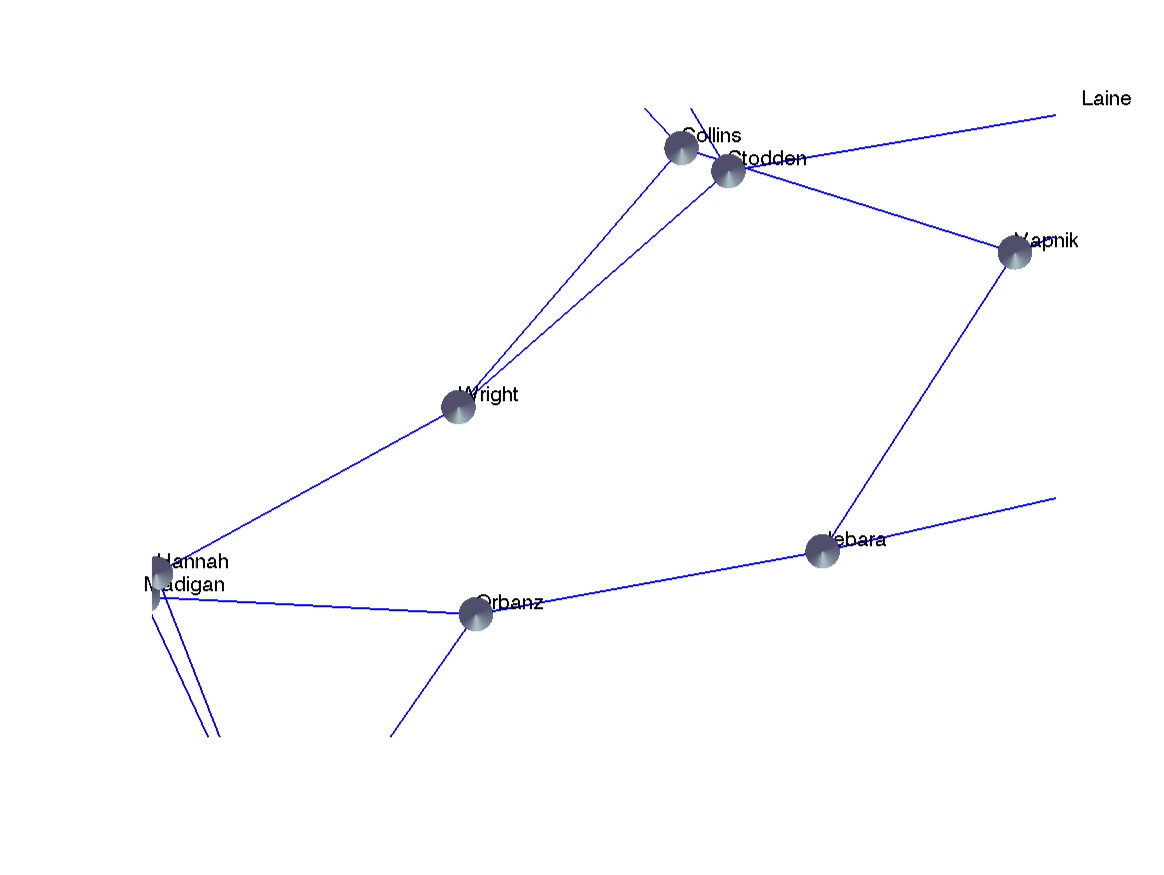}

\begin{center}
\textit{\footnotesize Figure 2. Sub-network of Figure 1 showing people
from Machine Learning, Statistics and Natural Language Processing }
\par\end{center}{\footnotesize \par}

\includegraphics[scale=0.58]{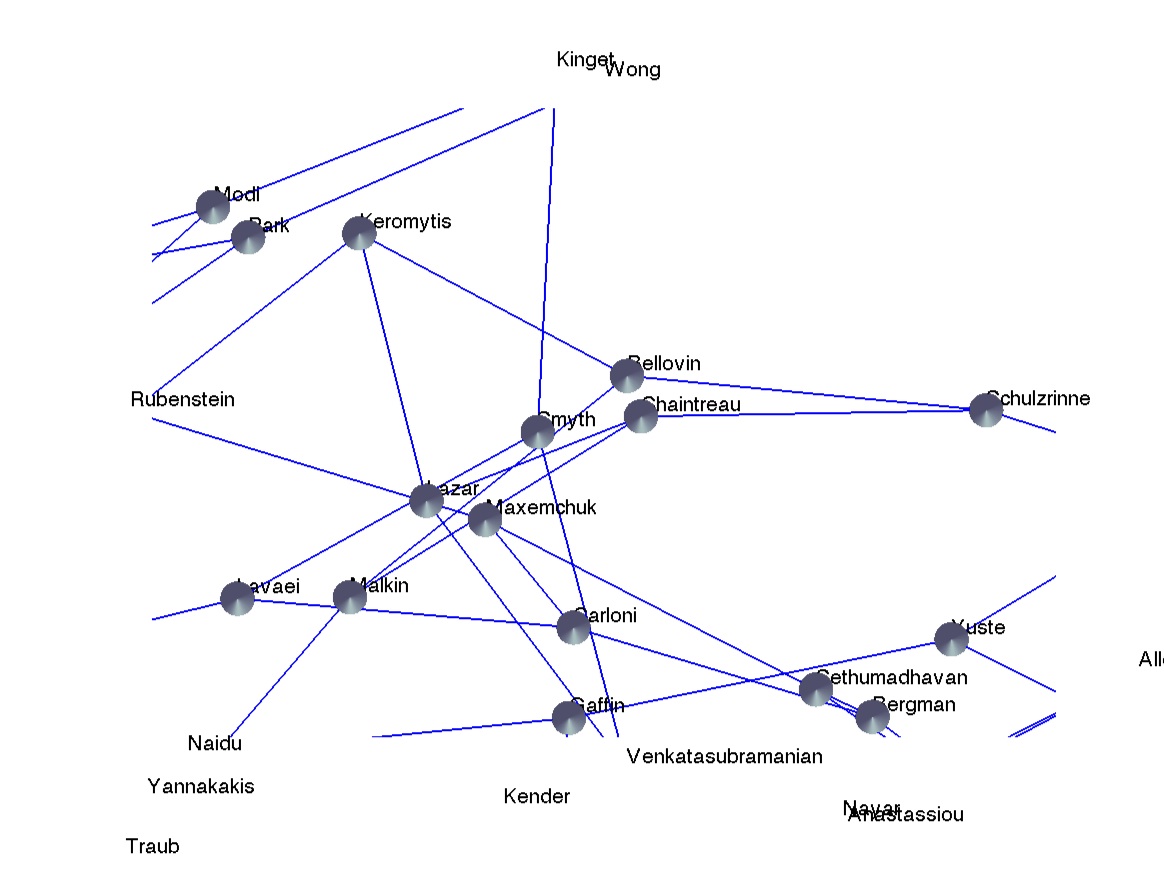}

\begin{center}
\textit{\footnotesize Figure 3. Sub-network of Figure 1. showing un-expected
connections}
\par\end{center}{\footnotesize \par}

The above figure depicts shows a few similarities which do not tally
with expected behavior. This can be explained by limiting the content
of the data just to the abstracts. Vital information pertaining to
the specific field of study of each author has more chance of appearning
more frequently if the entire paper is considered instead of just
the abstracts.\\
\\
\includegraphics[scale=0.75]{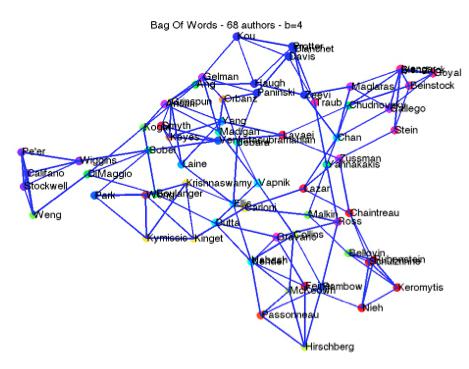}

\begin{center}
\textit{\footnotesize Figure 4. Aggregated Bag Of Words model - 68
authors - 10 papers each}
\par\end{center}{\footnotesize \par}

The above figure depicts the aggregated bag of words models. The color
assignments for each author represent the most frequent topic derived
from the topic distributions generated by the author topic model with
similar model parameters (with number of topics = 20).\\
\\
\includegraphics[scale=0.6]{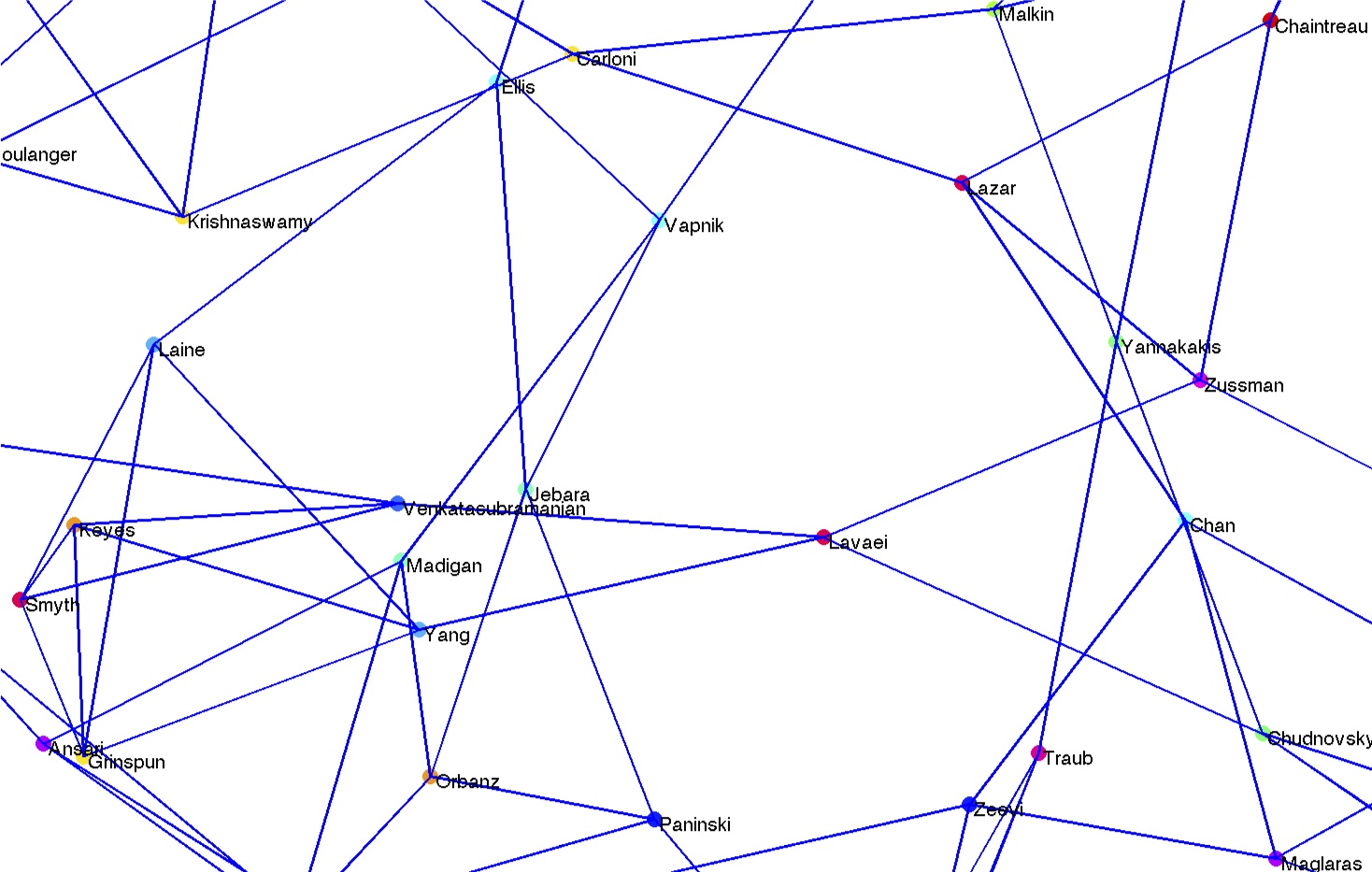}

\begin{center}
\textit{\footnotesize Figure 5. Magnified version of Aggregated Bag
of Words Model - 68 authors - 10 papers each}
\par\end{center}{\footnotesize \par}

The above figure depicts the magnified visualization for a different
dataset. The similarity between Machine Learners and Statisticians
is apparent in the figure. More similarities amiong other authors
can be observed in this figure as well \textbf{eg.} Complexity Theory,
Algorithms. Also we can see that this model is more often than not
in agreement with the A-T Model due to the high frequency of similar
colors of the nodes connected to each other.\\
\\
\includegraphics[scale=0.75]{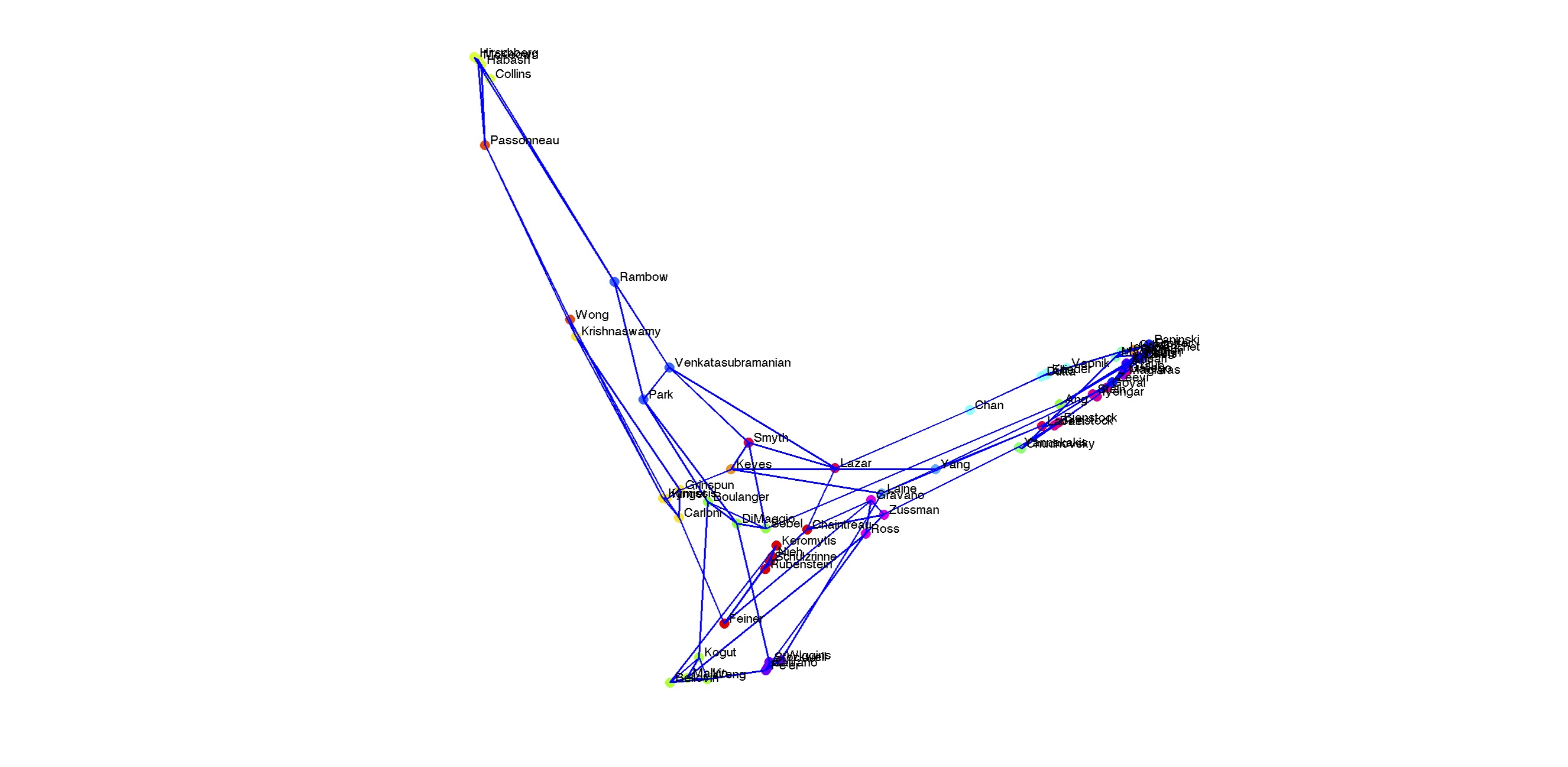}

\begin{center}
\textit{\footnotesize Figure 6. The A-T Model - 68 authors - 10 papers
each, 20 topics}
\par\end{center}{\footnotesize \par}

The above figure depicts the A-T model visualization.\\
\\
\includegraphics[clip,width=15cm,height=9cm]{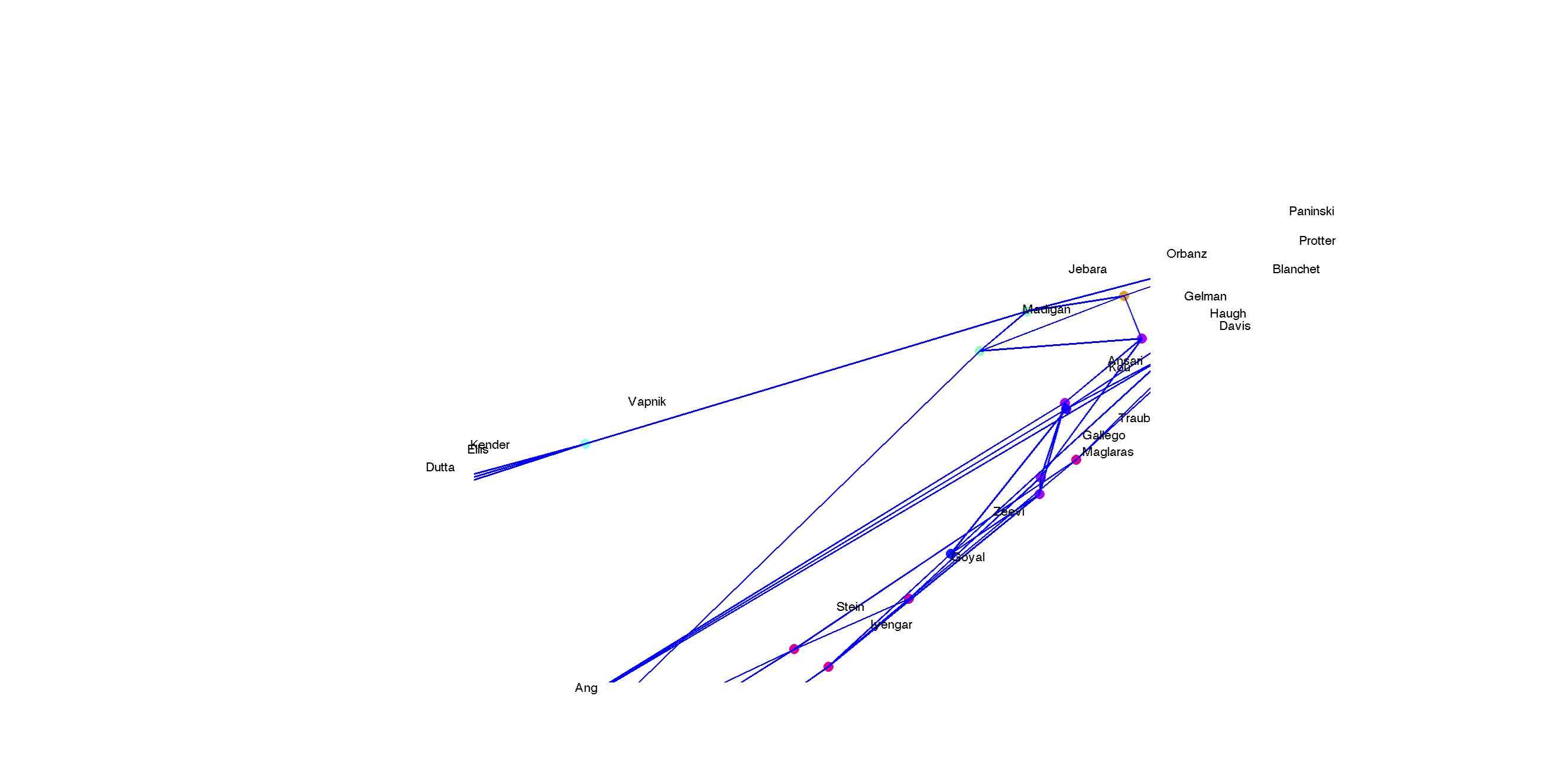}

\begin{center}
\textit{\footnotesize Figure 7. The A-T Model - Machine Learning -
Staistics - Neuroscience Sub-Network}
\par\end{center}{\footnotesize \par}

This ``subnetwork'' shows the inter connections between authors
of Machine Learning and Statistics although there are a few inconsistencies
in the topic allocations as can be see from the color of the labels.\\
\\
\includegraphics[scale=0.4]{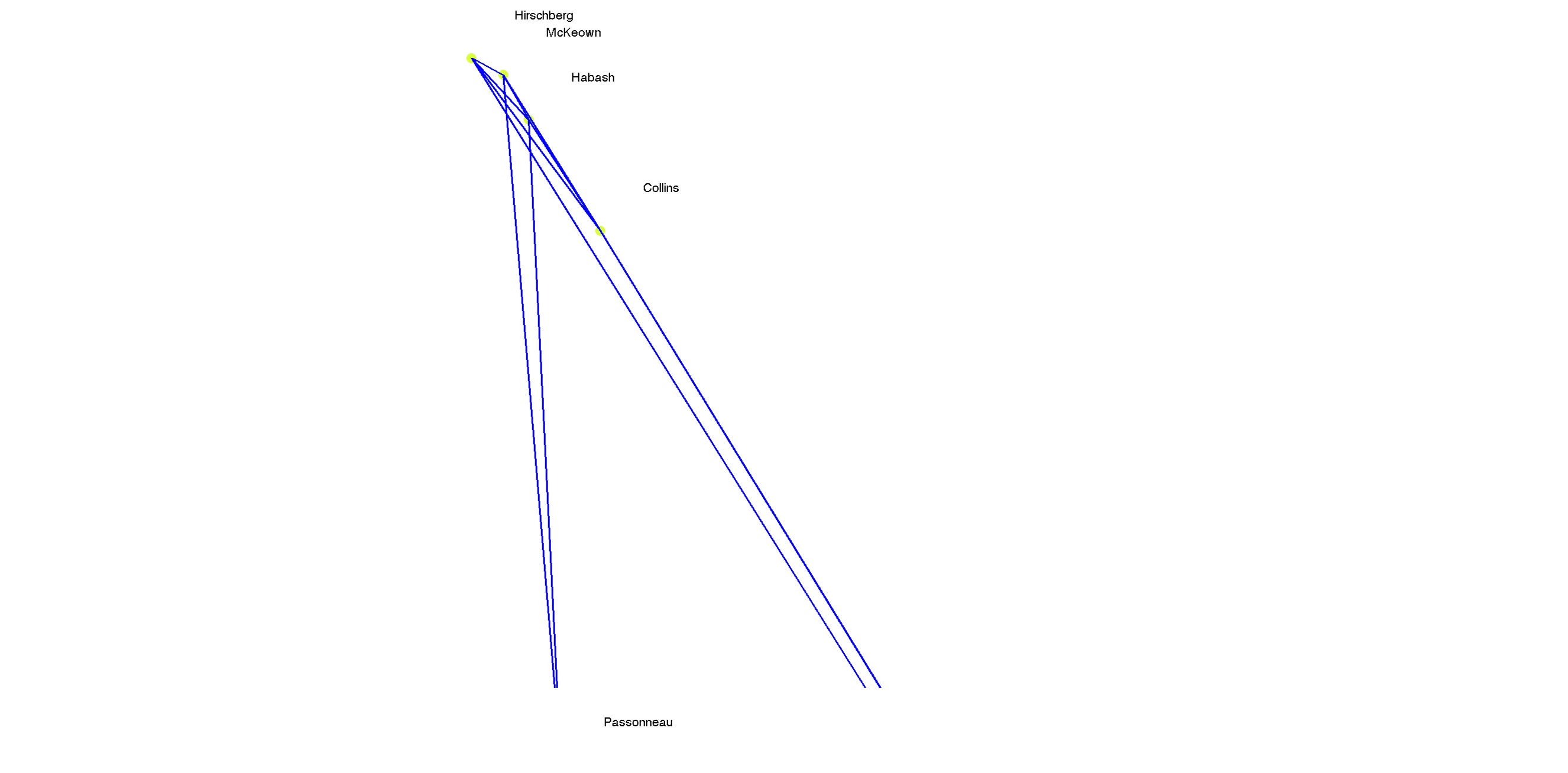}

\begin{center}
\textit{\footnotesize Figure 8. The A-T Model - Natural Language Processing
Sub-Network}
\par\end{center}{\footnotesize \par}

\includegraphics[scale=0.3]{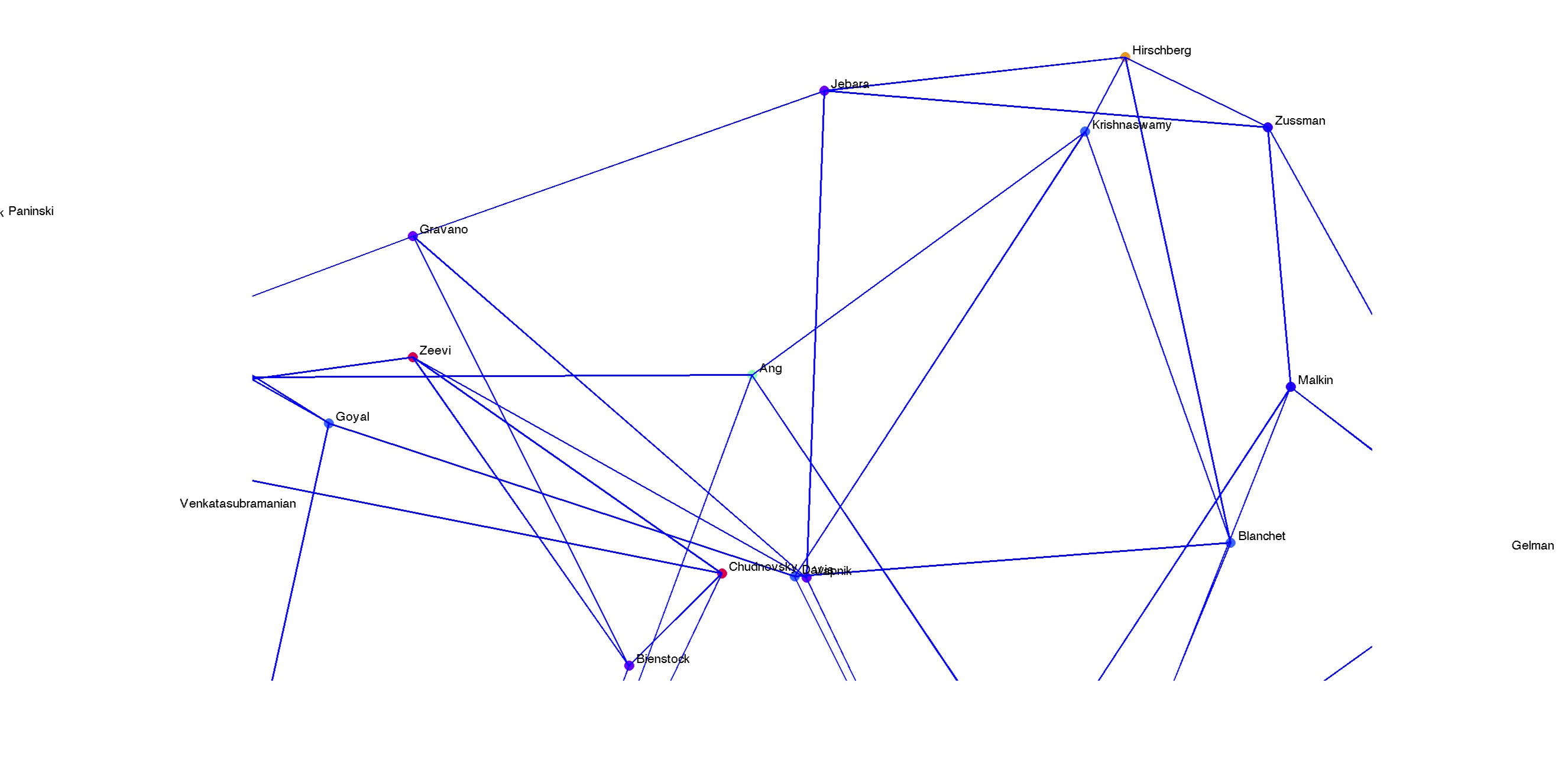}

\begin{center}
\textit{\footnotesize Figure 9. Aggregated LDA Model - 68 authors
- 10 papers each, 20 topics}
\par\end{center}{\footnotesize \par}

The above figure depicts the visualization using LDA to determine
the topic distributions of each document and aggregating over the
set of documents for each author to arrive at a pseudo-topic distribution.
The aggregated topic distributions do not tally with the expected
behavior- as was expected.

\section{Hierarchical A-T Model\label{sec:Hierarchical-Author-Topic-Model}}

From the A-T Model, a natural extension that is possible is to extend
it for data that can be naturally generated from a hierarchical categorical
model.

\subsection{Intuition}

If the data belonged to such a framework, using the information from
different layers of categories might help in distinguishing the data
more accurately than just using a single layer of topics as done by
the \textbf{A-T Model. }Any such discriminating feature will help
in achieving a more accurate similarity score between authors, which
can help improve the visualization. \\
 \\
Also the hierarchy of categories is prevalent in most data including
the author-papers dataset where each document / author can be segregated
starting from a broad range of topics (eg. departments, fields) to
the most specific (eg. Machine Learning, Complexity Theory). Also
various other datasets can be naturally classified into a hierarchy
of classes eg. Kingdom-Genus-Species ( Biology), Compounds in Organic
Chemistry etc. Thus this model can be used for visualizing datasets
conforming to this scenario.

\subsection{Model}

\includegraphics[scale=0.14]{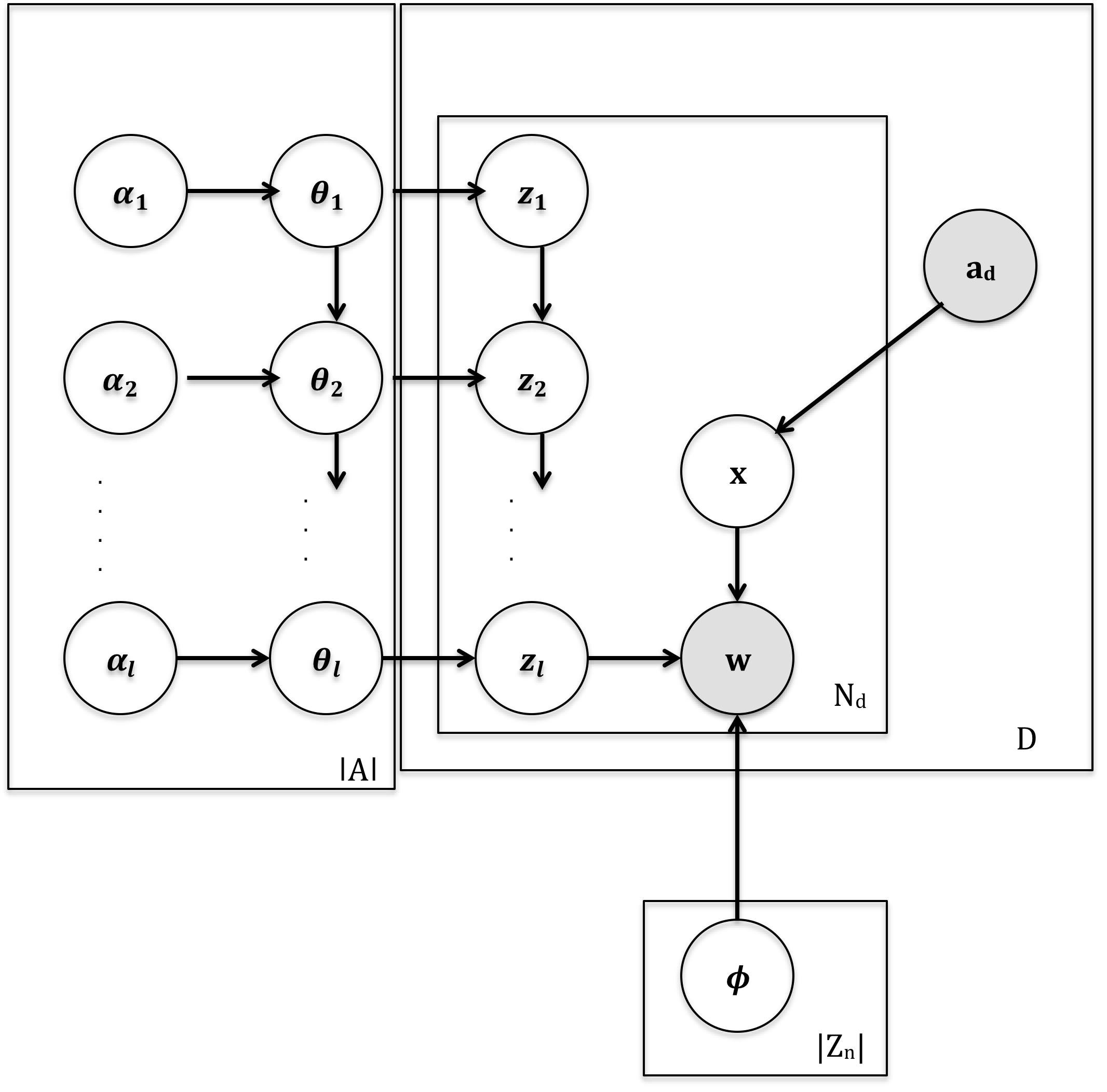}

\begin{center}
\textit{\footnotesize Figure 10. The Hierarchical A-T Graphical Model}
\par\end{center}{\footnotesize \par}

The graphical model presented above represents the hierarchical A-T
Model. Each of the $\alpha_{i}'s$ are the Dirichlet priors of the
$\theta_{i}'s$ which correspond to the multivariate distribution
at the $i^{th}$ level. Each of the $z_{i}'s$ are the $i^{th}$ level
topic assignments, with $z_{l}$ corresponding to topic assignment
at the leaf level (\textbf{i.e. }the most specific topic assignment).
These leaf topics being the most specific are part of the word-topic
probability matrix $\phi$ . Each of the $z_{i}'s$ depend on $z_{i-1}$.

\subsection{Bayesian Inference}

Inference of the parameters $\theta$ and $\phi$ can be solved by
using E-M. But this paper uses the technique of collapsed Gibbs sampling
\textbf{{[}3{]}} to arrive at the expresssion $p(Z,X,A,W;\alpha,\beta)$
by integrating out the correspondin parameters. \\
\textbf{}\\
\textbf{Note: }The complete derivation is provided in the supplementary
paper provided.

\subsection{Visualization}

The dataset consisting of 68 authors with 10 papers for each author
was visualized using the Hierarchical AT-Model with 2 layers of hierarchy
(2 divisions at the root, 20 divisions at the leaf -> 20 topics in
total ). The visualizations are provided below:-

\includegraphics[scale=0.35]{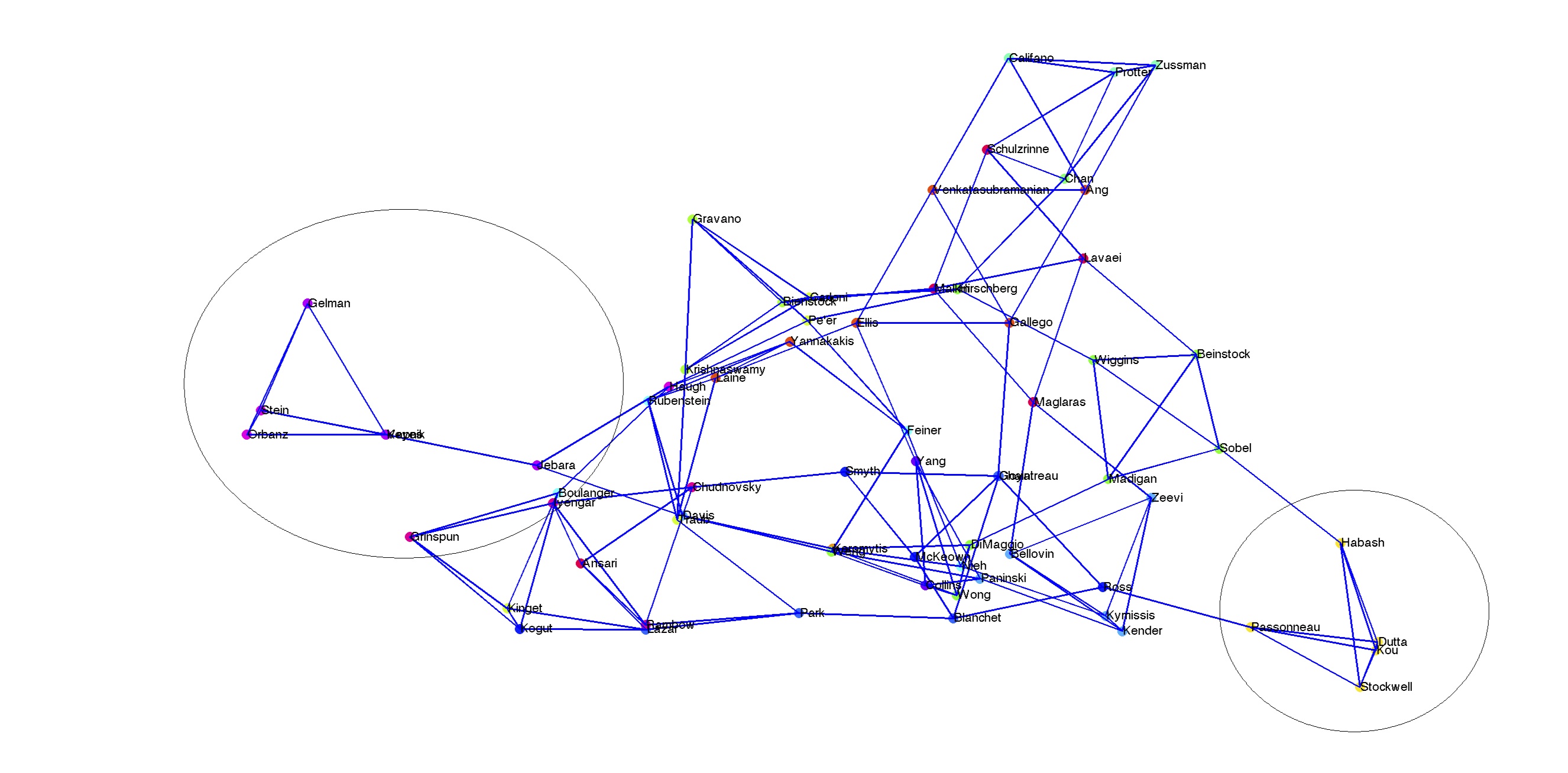}

\begin{center}
\textit{\textcolor{black}{\small Figure 11. Hierarchical AT-Model
68 authors - 10 papers each}}
\par\end{center}{\small \par}

Using the Hierarchical AT-Model on the above dataset gives a rather
unique visualization compared to the the visualization derived from
using the A-T Model. Two ``subnetworks'' circled in the above diagram,
show us that both relationships present in the previous visualizations
are preserved and also improved upon. One such sub-network is shown
in the next figure.\\

\includegraphics[scale=0.35]{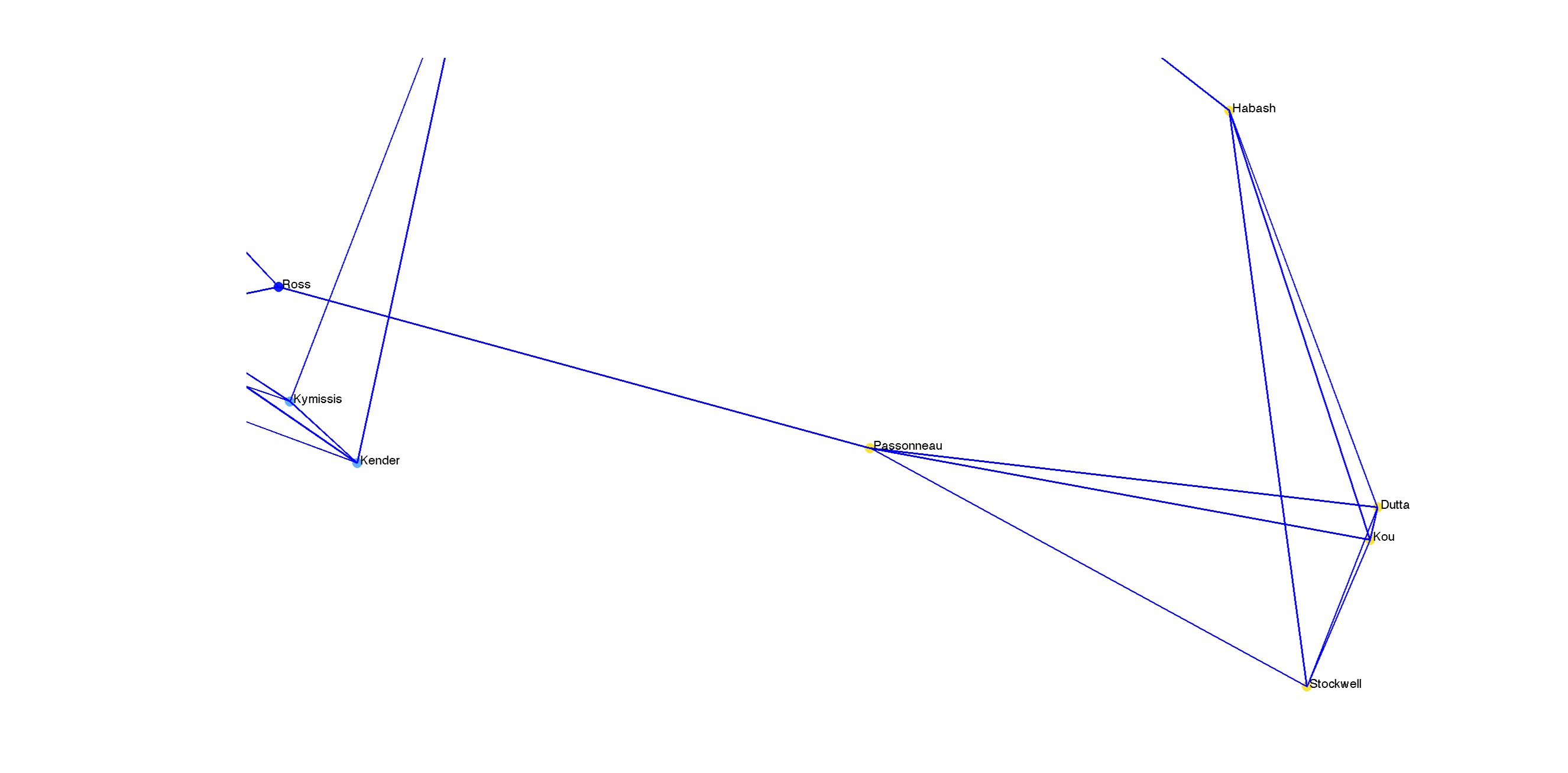}

\begin{center}
\textit{\textcolor{black}{\small Figure 12. CCLS - sub-network}}
\par\end{center}{\small \par}

One such improvement in the visualization is observed between the
authors whose primary focus is in the area of Natural Language Processing.
The authors who have primarily published from the CCLS (http://www.ccls.columbia.edu)
are more similar compared to others whose publications are not primarily
from CCLS, proving that more complex sub-relations between similar
entities can be captured by the HAT model.

\section{Conclusion}

Firstly, it was observed that using the full paper information for
each author over just the abstracts gave better visualization with
the \textbf{Aggregated Bag of Words Model}. This can be attributed
to higher frequency of certain domain-specific words in the papers
compared to the abstracts. Certain abstracts just constitute a sentence
while others might be equivalent to around 10 sentences. This might
overestimate certain word probabilities or might lead to sparse word
probability vectors. \\
\\
It can be observed that the \textbf{A-T Model }considerably provides
better visualization than the Aggregated Bag Of Words Model. This
can be attributed to the hidden topic information inferred from the
data which provides more information about the class to which the
authors belong facilitating more accurate similarity score measurements.
Also using \textbf{LDA} to infer the topic distributions of the documents
and aggregating over the authors did not provide a reasonable visualization
as was expected generally. \\
\\
The visualization generated by the \textbf{Hierarchical A-T Model}
seems to be on par with the A-T Model in preserving the relationships,
but in a few cases provides an improvement by taking advantage of
inherent hierarchy of classifications by providing more consistent
and accurate topic assignments. Also in the case of authors-papers
dataset, we can represent each author with topic distribution from
different layers, giving rise to different kernels (including different
combinations of two or more individual layer kernels) for computing
the similarity score w.r.t \textbf{MVE}. \\
\\
Logistically, to achieve an improvement on the visualization from
the A-T Model, the right set of parameters have to be tuned into the
model in particular the number of layers in the hierarchy of classification
of the data as well as the number of divisions at each layer. In case
of the author-paper dataset, it can be inferred logically that 2 layers
of hierarchy are required - the first layer broadly categorizes topics
based on department (\textbf{e.g. }Engineering, Physics, Biology,
Law, Business etc.), and the second layer provides a more specific
classification conditioned on the previous layer's topic. In practice
the number of divisions of the top layer pertaining to the author-paper
dataset can be expected to lie between 2 and 6. The number of leaf
topics is decided via cross-validation as is the case for A-T Model.\\
\\
In cases where just a single layer of topics exist, the Hierarchical
Model can handle this case by setting the number of divisions at the
root to be equal to 1 and the number of layers to 2 with the same
number of leap topics. The resulting visualizations were also found
to be equivalent for the dataset with 68 authors - 10 papers as well
further supporting the credibility of the Hierarchical A-T Model.

\section*{Future Work}

The dataset consisting of 68 authors - 10 papers each was chosen due
to convenience - Validating the visualization with the expected behavior
is easier since the authors of this paper are aware of the research
interests and works of the authors of this dataset. More datasets
belonging to different domains will be used to further test the merits
of the proposed model \textbf{e.g. }A set of 2037 authors with 1740
papers in total from the NIPS archives (1987-1999). Moreover we can
also experiment with certain modifications to the existing graphical
model in addition to modifying the kernel based on different layers.

\end{document}